\documentclass[showpacs,twocolumn,floatfix,superscriptaddress]{revtex4}
\usepackage{graphicx}
\usepackage{amsmath}
\newcommand{\beq}{\begin{equation}}
\newcommand{\eeq}{\end{equation}}
\newcommand{\bea}{\begin{eqnarray}}
\newcommand{\eea}{\end{eqnarray}}
\newcommand{\pdag}{{\phantom{\dagger}}}

\begin{document}
\title{Correlations of spin currents through a  quantum dot induced by the Kondo effect  }
\author{M.\ Kindermann}
\affiliation{ Department of Physics, Massachusetts Institute of Technology,
Cambridge MA 02139, USA}
\date{May 2004}
\begin{abstract}
  We study correlations of spin currents flowing through a Coulomb blockaded quantum dot. While vanishing for elastic co-tunneling,  these correlations develop as the quantum dot enters the Kondo regime. They are a manifestation  of Kondo physics in  quantum dots. We  demonstrate that the spin current correlator is non-perturbative in the Kondo coupling.  
\end{abstract}
\maketitle

Measurements of current correlations in electrical conductors give valuable information about the microscopic interactions. An important example are shot noise measurements on  conductors in the fractional quantum Hall regime \cite{Hei97}. They have demonstrated that the excitations of those interacting systems carry fractions of the elementary charge.    Another well-studied interaction phenomenon in nanostructures is the Kondo effect \cite{Kondo-popular}. It is usually observed as an increase of the conductance through a Coulomb blockaded quantum dot (QD) upon lowering its temperature \cite{kondo}. Its microscopic origin are ground-state correlations between an  unpaired spin on the QD and the spins of electrons in the adjacent leads. This suggests that a natural way to reveal  Kondo correlations should be through spin-dependent quantities.  In this letter we show that indeed Kondo physics manifests itself in  correlations of  spin currents through a QD.
 The measurement of  spin currents is one of the goals of the rapidly developing field of spintronics \cite{Aws01}. The application discussed  here requires  a measurement method that does not disrupt the Kondo effect. In particular it must not spin-polarize the leads. One possibility based on the coupling of moving spins to an electric field has been put forward in \cite{DiL03}.

 We  study the correlator
\begin{equation} \label{eq:Cud}
C_{\uparrow\downarrow} = \int{dt\, \langle I_{\uparrow}(0) I_{\downarrow}(t) \rangle - \langle I_{\uparrow}\rangle\langle I_{\downarrow} \rangle }
\end{equation}
between the currents $I_{\uparrow}$ and  $ I_{\downarrow}$ of the numbers of  spin-up and spin-down electrons flowing  through a QD.  We assume that the applied bias voltage $V$ and the temperature $T$ are well below  level spacing $\epsilon_d$ and charging energy $U$ of the dot, $eV,kT \ll \epsilon_d,U$ (we set $\hbar=k=1$). The system is then well described by the Anderson single-level impurity model
\begin{eqnarray}
H_A &=&H_L^{(0)} + H_D  + U n_{d \uparrow} n_{d \downarrow}  
+ \sum_{k \beta \sigma} v_{\beta} (d^\dagger_{\sigma} c^\pdag_{k\beta\sigma} + h.c.) \nonumber \\
 H_L^{(0)} &=&  \sum_{k=-\Lambda^{(0)}\atop \beta, \sigma}^{\Lambda^{(0)}} ( \epsilon_{k}-\mu_{\beta}) c^\dagger_{k \beta \sigma} c^\pdag_{k \beta \sigma} ,\;\; H_D=- \sum_\sigma \epsilon_{d}  n_{d\sigma} . \nonumber \\
\end{eqnarray}
$k$ labels the electrons' momentum, $\sigma\in \{\uparrow,\downarrow\}$ their spin and $\beta\in\{L,R\}$ the lead (left or right of the QD),  $ n_{d\sigma}=d^\dagger_{\sigma} d^\pdag_{\sigma}$. $\mu_{\beta}$ is the electrochemical potential of lead $\beta$. We have $\mu_L-\mu_R=eV$ and  we choose the zero of energy at the Fermi level.  If the dot is weakly coupled to the leads, $ \Gamma_{\beta} \ll U-\epsilon_d, \epsilon_d$ ($\Gamma_{\beta}=2 \pi \nu |v_{\beta}|^2$ and $\nu$ is the density of states in the leads), and the QD is operated in a Coulomb blockade valley, charge fluctuations on the dot level $d$ are strongly suppressed. If additionally  $d$ is  occupied by a single electron, the spin of that electron is the only relevant degree of freedom of the QD.    One obtains the corresponding low-energy Hamiltonian by a Schrieffer-Wolff transformation \cite{Hewson},
\begin{eqnarray} \label{eq:HK}
H^{(0)}&=& H^{(0)}_L + H^{(0)}_K, \nonumber \\
H^{(0)}_K &=&  \sum_{k,\beta,\sigma \atop \tilde{k}, \tilde{\beta}, \tilde{\sigma}}  c^{\dagger}_{k \beta\sigma} c^\pdag_{\tilde{k} \tilde{\beta} \tilde{\sigma}} \left( J_{\beta, \tilde{\beta}}^{(0)}  \hat{S}_a s^a_{\sigma \tilde{\sigma}} + \frac{\tilde{J}_{\beta,\tilde{\beta}}^{(0)}}{4} \delta_{\sigma\tilde{\sigma}}\right).
\end{eqnarray}
$J^{(0)}$ are the amplitudes of scattering processes of lead electrons that involve the spin $S$ of the QD,  whereas $\tilde{J}^{(0)}$ accounts for regular potential scattering. $s^a=\sigma^a/2$, where $\sigma^a$ are Pauli matrices. We assume that  either $ \epsilon_d \ll U-\epsilon_d$ or  $ \epsilon_d \gg U-\epsilon_d$, such that $|\tilde{J}^{(0)}|=|J^{(0)}|$.  The relative sign depends on whether  transport occurs by emptying the level $d$, that is   $ \epsilon_d \ll U-\epsilon_d$ ($\tilde{J}^{(0)}=J^{(0)}$), or by doubly occupying it, that is  $ \epsilon_d \gg U-\epsilon_d$ ($\tilde{J}^{(0)}=-J^{(0)}$).

\begin{figure}
\includegraphics[width=7cm]{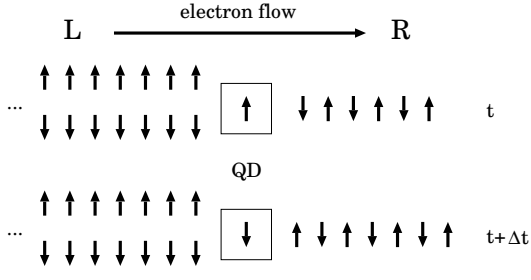}
\caption{ Illustration of the mechanism that introduces spin-current correlations in the Kondo regime. Only the relevant spin-flip scattering processes  are depicted. The QD has different spin at consecutive scattering times  $t$ and  $t+\triangle t$. Consequently, the outgoing electrons carry alternating spin.    }  \label{diagram} 
\end{figure}

Before giving the details of the calculation of  $C_{\uparrow\downarrow}$ we  motivate its results. For this we focus on the case $\tilde{J}^{(0)}=-J^{(0)}$.
In the co-tunneling regime, when only processes to lowest order in $J$ are relevant, transport occurs then  through virtual states with a doubly occupied level $d$. These virtual  states decay into  states with one electron of either spin on the QD, contributing to $ I_{\uparrow}$ or $ I_{\downarrow}$ with equal probabilities.   To lowest order in $J$ there are therefore no  correlations between  $ I_{\uparrow}$ and $ I_{\downarrow}$,   $C_{\uparrow\downarrow}=0$. It is important to note, that if $S=\sigma$ co-tunneling does not allow for the transfer of an electron  with spin $\sigma$  without  flipping  $S$ because of the Pauli principle. This changes at low temperatures, in the Kondo regime. Higher order tunneling processes  can then transfer  electrons with either spin  without flipping $S$. Being equally likely, these ``non-spin-flip'' processes do not contribute to  $C_{\uparrow\downarrow}$.  The spin current produced by  processes that flip $S$, however, is correlated with the state of $S$.  Whenever the spin on the dot is  ``up'', it can flip down and make a contribution to   $ I_{\uparrow}$. If $S$ is ``down'', a spin-flip process produces a spin-down current. Since in the absence of external spin-relaxation processes every flipping of $S$ to ``up'' has to be followed by a spin-flip to  ``down'', both processes  occur alternatingly and  equally often, as illustrated in Fig.\ \ref{diagram}. This introduces correlations between   $ I_{\uparrow}$ and $ I_{\downarrow}$   and we expect  $C_{\uparrow\downarrow}$ to be non-vanishing in the Kondo regime. Formally, the Pauli blocking in the co-tunneling regime is described by  Hamiltonian (\ref{eq:HK})  through an  interference of the amplitudes $J^{(0)}$ and $\tilde{J}^{(0)}$. It is lifted in the Kondo regime that is described  by an effective  Hamiltonian of the form (\ref{eq:HK}) with   $J^{(0)} \gg \tilde{J}^{(0)}$.

We study the limit of a weak Kondo effect with small effective low-energy couplings $\nu J\ll 1$. One might expect that  perturbation theory in $J$ is then adequate to calculate $C_{\uparrow\downarrow}$. This perturbation expansion is, however, plagued by divergences due to the fact that the correlation functions of $S$ do not decay in time to low order in $J$. This indicates that  long-time spin  correlations have to be treated   non-perturbatively.    For this we use a method that is close in spirit  to the master equation approach to current correlations in sequential tunneling  by Bagrets and Nazarov \cite{Bag02}.
 Since we are interested in the co-tunneling and the  Kondo regime we cannot directly apply their method and we  derive a variant of it.  Our approach is quantum mechanical  at short time scales while  non-perturbative and classical  at times that exceed the decoherence time of $S$. 
 
In the limit of interest $\nu J \ll 1$,  we can  apply the perturbative renormalization group \cite{And70} to capture the short-time spin dynamics. In \cite{Kam00} this approach  has been extended to   non-equilibrium situations as considered here.  Following  \cite{Kam00}  we eliminate virtual transitions into high energy electron states in the leading logarithm approximation down to a scale $\Omega \gtrsim eV$. This results in an effective  Hamiltonian $H$ that has the form of Eq.\  (\ref{eq:HK}), but renormalized coupling constants $J$ and $\tilde{J}$ instead of $J^{(0)}$ and $\tilde{J}^{(0)}$ and reduced bandwidth $\Lambda$ (instead of $\Lambda^{(0)}$) with $\epsilon(\Lambda)=\Omega$.   Spin fluctuations  enhance the spin-flip amplitude  to $J=1/2\nu \ln(\Omega/T_K)$, where $T_K$ is the Kondo temperature, while  $\tilde{J}=\tilde{J}^{(0)}$. Our limit $\nu J \ll 1$ implies  $eV \gg T_K$. We additionally assume that the temperature is smaller but of the same order of magnitude as the voltage, $eV \gtrsim T$.

The central  object in our approach is the generating function of  spin-current correlators
\beq  \label{eq:Z}
 {\cal Z}(\lambda)={\rm Tr}\, e^{-i \lambda_{\beta \sigma} N^{\beta \sigma} }\, e^{-iHt}\, e^{i\ \lambda_{\beta \sigma} N^{\beta \sigma} } \, \rho^{\rm (in)}\, e^{iHt}.
 \eeq
 $\rho^{\rm (in)}$ is the initial density matrix of the conductor and ${\cal Z}$  generates
 moments of the number $\triangle N^{\beta\sigma}$ of spin $\sigma$ electrons that is transferred into lead $\beta$ during time $t$,
  \begin{equation} \label{eq:genZ}
 \left\langle  \prod_{\beta,\sigma} \left(\triangle N^{\beta\sigma}\right)^{p_{\beta\sigma}} \right\rangle =  {\cal Z}^{-1} \prod_{\beta,\sigma} \left(i \frac{\partial}{\partial \lambda_{\beta\sigma}} \right)^{p_{\beta\sigma}}  {\cal Z}(\lambda)\Big|_{\lambda=0}.
 \end{equation}
 ${\cal Z}$ as defined in  Eq.\  (\ref{eq:Z}) is a Keldysh  partition function. Its two time development operators can be implemented by operators on the two branches of the Keldysh time-contour \cite{Naz99}.
The exponentials $\exp\{\pm i \lambda_{\beta \sigma}N^{\beta\sigma} \}$ 
can be absorbed into phases for the tunneling terms in $H$. By inserting complete sets of Fermion coherent states \cite{Neg88} we then derive a mixed representation  \cite{Naz01} of ${\cal Z}$ as a path integral over a time-ordered spin operator expression,
\beq \label{eq:path}
  {\cal Z}(\lambda)= {\rm Tr}_S\, T_c\,  \int{ {\cal D} {c}^* {\cal D} c  \, \rho_S^{\rm (in)} \, e^{-i {c}^*({\cal G}^{-1}+{\cal J}^{\lambda}) c}}.
  \eeq
The Fermion fields $c$ carry  Keldysh, lead,  spin, momentum, and frequency indices $\alpha$, $\beta$, $\sigma$, $k$, and $\omega$. ${\cal G}$,  a matrix in this space, is the electron Green function corresponding to $H_{L}$.   ${\cal J}^{\lambda}$ describes tunneling processes and contains spin operators. $T_c$ denotes operator ordering along the Keldysh time-contour and relative to 
 the initial spin density matrix  $\rho_S^{\rm (in)}$. The trace ${\rm Tr}_S$ over spin operators is taken. We have
\begin{widetext}
\beq
 {\cal J}^{\lambda}_{\alpha\beta\sigma k\omega,\alpha'\beta'\sigma'  k'\omega'} =  \tau^z_{\alpha\alpha'} \left(J_{\beta\beta'} s^a_{
\sigma,\sigma'} \hat{S}_{a}^{\alpha,\omega\omega'} +\frac{\tilde{J}_{\beta\beta'}}{4} \delta_{\sigma\sigma'}\delta_{\omega\omega'}\right) e^{-i(\lambda_{\beta \sigma}-\lambda_{\beta' \sigma'}) \tau^z_{\alpha\alpha'}/2}
 \eeq
with  the third Pauli matrix  $\tau^z$ and spin operators $ \hat{S}_{a}^{\alpha,\omega\omega'}= \int{dt\, e^{-i(\omega-\omega')t}  \hat{S}_{a}^{\alpha}(t)}$  on the Keldysh branch $\alpha$. Integrating over the  Fermions in  Eq.\ (\ref{eq:path}) we obtain
 \beq \label{eq:Zla}
 {\cal Z}(\lambda) =   {\rm Tr}_S\,  T_c \,\rho^{\rm (in)}_S\,  e^{{\rm Tr}\, \ln(1 + {\cal G} {\cal J}^{\lambda})}\,{\cal Z}|_{J=\tilde{J}=0} =  {\rm Tr}_S\,  T_c \,\rho^{\rm (in)}_S \, e^{{\rm Tr}\,[ {\cal G} {\cal J}^{\lambda} -  \frac{1}{2} {\cal G} {\cal J}^{\lambda} {\cal G} {\cal J}^{\lambda} + {\cal O}(J^3)]} \equiv   {\rm Tr}_S\,  T_c \,\rho_S^{\rm (in)} \, e^{-{\cal L}^{\lambda}}.
 \eeq
The ${\cal O}(J)$ term of ${\cal L}^{\lambda}$  does not contain spin operators, while the terms of ${\cal O}(J^3)$ are  negligible for $\nu J \ll 1$. To save space we evaluate here only the ${\cal O}(J^2)$ term  ${\cal L}^{\lambda}_o$  that derives from the spin off-diagonal matrix elements of  ${\cal J}^{\lambda}$,
\beq  \label{eq:Sud}
 {\cal L}^{\lambda}_o =  \sum_{k,k'\atop\alpha\alpha '\beta\beta'}\frac{|J_{\beta\beta'}|^2}{4} \tau^z _{\alpha \alpha }\tau^z_{\alpha '\alpha'}  e^{i(\lambda_{\beta\uparrow}-\lambda_{\beta'\downarrow}) (\tau^z_{\alpha \alpha }-\tau^z_{\alpha '\alpha '})/2}\int{dt\,dt'\, G^{\beta k}_{\alpha,\alpha'}(t'-t)  G^{\beta' k'}_{\alpha',\alpha}(t-t')  \hat{S}_-^{\alpha'}(t) \hat{S}_+^{\alpha}(t')},
\eeq 
\end{widetext}
where $G^{\beta k}_{\alpha,\alpha'}(t) = \int{(d\omega/2\pi)\, e^{-i\omega t}  {\cal G}_{\alpha\beta\uparrow k\omega,\alpha'\beta\uparrow  k\omega}} $ and $\hat{S}^{\alpha}_{\pm}=\hat{S}^{\alpha}_x\pm i \hat{S}^{\alpha}_y$.
The Green functions in Eq.\ (\ref{eq:Sud}) summed over $k,k'$ decay exponentially over the time  $\tau_T=1/T$. At time scales longer than $\tau_T$ ${\cal L}^{\lambda}_o$ is therefore local in time. It moreover couples spin-flip processes occurring at the same time  on different branches of the Keldysh contour. This leads to  classical behavior at long times and it allows for a description  along the lines of \cite{Bag02}. One could  obtain the effective theory on the scale $\tau_T$ by integrating out fast spin fluctuations in a path integral for the original model Eq.\ (\ref{eq:HK}) \cite{Ham70}. Equivalently we integrated out high frequency spin fluctuations   in the Hamiltonian formalism. Due to the form of the Green functions in the effective model $H$ with reduced bandwidth, 
 $ {\cal L}^{\lambda}$ is then strongly suppressed for frequencies  $\omega > \Omega$ and these fluctuations  contribute to ${\cal Z}$ only negligibly.   The corrections to the scaling logarithms due to spin fluctuations in the frequency range $T<\omega<\Omega$ are for our choice of parameters $\Omega \gtrsim eV\gtrsim T$ of ${\cal O}(1)$ and  to leading order in the logarithms negligible as well. To this accuracy $ {\cal L}^{\lambda}_o$  therefore equals  the corresponding piece of the  effective theory at the scale $\tau_T$. The contribution due to the diagonal elements of  ${\cal J}^{\lambda}$  has an analogous structure and we conclude that ${\cal L}^{\lambda}$ is local  on the time scale $\tau_T$.  
In Eq.\ (\ref{eq:Zla}) we have expressed ${\cal Z}$ as the trace over a time dependent density matrix $\rho_S^{\lambda}(t)=T_c \,\rho_S^{\rm (in)} \, \exp({-{\cal L}^{\lambda}})$.
 Because of the locality of  $ {\cal L}^{\lambda}$, $\rho_S^{\lambda}$ obeys   an ordinary differential equation. The  off-diagonal entries of $\rho^{\lambda}_S$ decay exponentially under evolution with that equation. This shows that  the spin dynamics is indeed classical on the time scale $\tau_T$.  It is therefore sufficient to study the evolution of the diagonal elements $p_{\uparrow}^{\lambda}$ and $p_{\downarrow}^{\lambda}$ of $\rho_S^{\lambda}$, that we collect into a vector $p^{\lambda}=(p_{\uparrow}^{\lambda},p_{\downarrow}^{\lambda})$.  
$p^{\lambda}$ obeys
\beq \label{eq:diff}
 \partial_t p^{\lambda} = - \hat{L}^{\lambda} p^{\lambda}, 
\eeq
\begin{widetext}
\beq \label{eq:L}
 \hat{L}^{\lambda}= \left( \begin{array}{cc}     \Gamma_S - \sum_{\beta\beta'} a_{\beta\beta '}\left(\left|\frac{J+  \tilde{J}}{2}\right|^2 C^{\lambda}_{\beta\uparrow,\beta '\uparrow} + \left|\frac{J-  \tilde{J}}{2}\right|^2  C^{\lambda}_{\beta\downarrow,\beta '\downarrow}  \right)  &   -\Gamma_S-  \sum_{\beta\beta'} a_{\beta\beta '} |J|^2 C^{\lambda}_{\beta\downarrow,\beta '\uparrow} \\ -\Gamma_S -   \sum_{\beta\beta'} a_{\beta\beta '} |J|^2 C^{\lambda}_{\beta\uparrow,\beta '\downarrow} &    \Gamma_S- \sum_{\beta\beta'}   a_{\beta\beta '}\left(\left|\frac{J+  \tilde{J}}{2}\right|^2 C^{\lambda}_{\beta\downarrow,\beta '\downarrow} + \left|\frac{J-  \tilde{J}}{2}\right|^2  C^{\lambda}_{\beta\uparrow,\beta '\uparrow}  \right) \end{array} \right),
\eeq
\end{widetext}
\bea
C^{\lambda}_{\beta\sigma,\beta '\sigma '}&=&e^{-i(\lambda_{\beta\sigma}-\lambda_{\beta ' \sigma '})} -1, \nonumber \\
a_{\beta\beta '}&=& (\pi \nu)^2 \int{\frac{d\epsilon}{2\pi}[1-f_{\beta }(\epsilon)] f_{\beta'}(\epsilon)} .
\eea
 For conciseness we assume from now on that the QD is symmetrically coupled  to the leads, $J_{\beta\beta'}=J$, $\tilde{J}_{\beta\beta'}=\tilde{J}$.
$a_{\beta\beta'} |J|^2  $ are the usual rates for tunneling between two leads.
$\Gamma_S=  \sum_{\beta\beta '}  a_{\beta\beta '} |J|^2  + \gamma_S$ is the relaxation  rate of $S$. The first term accounts for spin-flips by conduction electrons. The second term $ \gamma_S$  has been   introduced  phenomenologically. It accounts  for  spin-flip processes that are not described by our model.  We assume $\gamma_S \ll T$, such that this additional spin-decoherence does not affect  Kondo correlations.
Eqs.\ (\ref{eq:diff}) with (\ref{eq:L}) differ from  their counterparts for sequential tunneling \cite{Bag02}  mainly  in how the electron state occupation numbers enter. Kondo correlations are accounted for by the renormalization of $J$. The off-diagonal entries of $\hat{L}^{\lambda}$ describe spin-flips of $S$. Because electrons can also be transferred without flipping $S$,  the diagonal elements of  $\hat{L}^{\lambda}$ contain counting factors $C^{\lambda}$ as well.

\begin{figure}
\includegraphics[width=6cm]{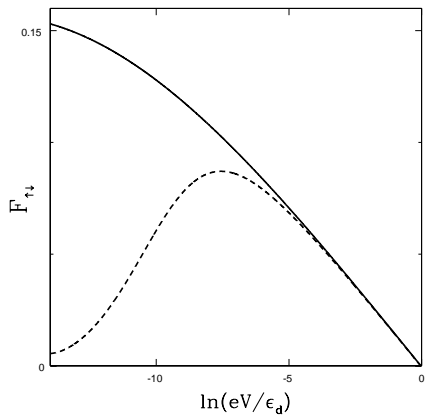}
\caption{Voltage dependence of the Fano factor $F_{\uparrow\downarrow}$ for $T=eV/30$, $\epsilon_d=1\,{\rm meV}$,  and $ T_K= 10^{-7} \,\epsilon_d$ (solid line: $\gamma_S=0$, dashed line:  $\gamma_S= 10^5\, {\rm s}^{-1}$). $F_{\uparrow\downarrow}$ develops in parallel  with Kondo correlations upon lowering the voltage/temperature.
 }  \label{Cud} 
\end{figure}

We integrate Eq.\ (\ref{eq:diff}) to obtain ${\cal Z}$. At long times $t$ it simplifies to the exponential of   the smallest eigenvalue of $-t \hat{L}^{\lambda}$.
$C_{\uparrow\downarrow}$ can then be obtained using  Eq.\ (\ref{eq:genZ}). We introduce transmission probabilities $\tau=(2\pi \nu J)^2$, $\tilde{\tau}=(2\pi \nu \tilde{J})^2$ and the spin-off-diagonal Fano-factor $F_{\uparrow\downarrow}=C_{\uparrow\downarrow}/I$ ($I$ is the mean current through the QD).  Although not within the limits of applicability of our theory, it is instructive to first take the zero temperature limit
\beq \label{eq:Fud0}
 F_{\uparrow\downarrow}= \frac{eV\tau}{2(3\tau+\tilde{\tau})(eV\tau+8\pi\gamma_S)}\, (\tau-\tilde{\tau}) .
 \eeq
  Eq.\ (\ref{eq:Fud0}) displays most clearly the main finding of this letter: While  $F_{\uparrow\downarrow}$ vanishes in the absence of Kondo correlations, when potential and spin-flip scattering processes have equal probabilities $\tau=\tilde{\tau}$, it grows non-zero in the Kondo regime.  We need to convince ourselves that this carries over to  finite temperature $T\simeq eV$.
  The first  correction to Eq.\ (\ref{eq:Fud0}) due to temperature, $\triangle F^{(1)}_{\uparrow\downarrow}= [(\tilde{\tau}-\tau)/(3\tau+\tilde{\tau})](T/eV)$ ($\gamma_S=0$) is reassuring: it again vanishes unless there are Kondo correlations. The full temperature dependence of   $F_{\uparrow\downarrow}$,
\begin{widetext}
\beq \label{eq:Fud}
 F_{\uparrow\downarrow}= \frac{ \tau v^2[\tau \cosh^2(v/2) - \tilde{\tau} \sinh^2(v/2)] -4\tau ( \tau + 8\pi \gamma_S/T)\sinh^2(v/2)}{(3\tau+\tilde{\tau})v[\tau v\sinh v + 4(\tau  + 4\pi\gamma_S/T) \sinh^2(v/2)]}, \;\;\; v=\frac{eV}{T},
 \eeq
 \end{widetext}
is shown in Fig.\ \ref{Cud}. In this plot both temperature and voltage are varied, at constant ratio $eV/T \gtrsim 1$. 
 Although the Fano factor   $F_{\uparrow\downarrow}$ at finite temperature does not reach its theoretical zero temperature maximum $1/6$, its qualitative behaviour is robust: it develops in parallel with  Kondo correlations. The effect is cut-off at small voltages $ eV \ll \gamma_S/\tau$ by external spin-flip processes. $S$ is then flipped randomly in between electron transfers and the mechanism illustrated in Fig.\ \ref{diagram} is inoperative.  For  Fig.\ \ref{Cud}  we have assumed typical experimental parameters  $\epsilon_d=1\, {\rm meV}$ and $\gamma_S= 10^5\, {\rm s}^{-1}$. Even longer spin relaxation times have been observed  \cite{Han03}.
 
We come back now to the  difficulties encountered with  perturbation theory in $J$. The zero temperature limit  Eq.\ (\ref{eq:Fud0}) shows most clearly that $C_{\uparrow\downarrow}$ in the model Eq.\ (\ref{eq:HK}) ($\gamma_S=0$) is  non-perturbative in $J$:  the limit  $\gamma_S\to 0$ implies that $\tau \gg \gamma_S/eV$, that is $J^2 \gg \gamma_S/\nu^2 eV$ and it  cannot be accessed in an expansion around $J=0$.   It  lies outside its radius of convergence.   Perturbation theory in $J$ misses correlations of scattering events over the time-scale $1/\Gamma_S$. These long-time correlations are manifest in  the frequency spectrum of $C_{\uparrow\downarrow}$. It can be obtained by a straightforward extension of our method to slowly time-dependent  $\lambda$. The approach is valid for frequencies $\omega \ll T$ and yields
\beq
C_{\uparrow\downarrow}(\omega)=\left(I F_{\uparrow\downarrow} + \frac{\tau T}{8\pi}\right) \frac{ \Gamma^2_S}{\omega^2/4+  \Gamma_S^2} -   \frac{\tau T}{8\pi}.
\eeq
The dispersion of $C_{\uparrow\downarrow}$ on the  scale $\Gamma_S$, that is of the order of    the mean current $I/e$  at small $\gamma_S$, is a consequence of the time-ordering of  spin-flips described above.

In conclusion, we have studied  correlations of spin currents  through a QD. We have found that these correlations can be induced by Kondo fluctuations on the QD. They are a new manifestation of the Kondo effect in QDs.

I thank J.\ von  Delft, L.\ I.\  Glazman, W.\ Hofstetter, L.\ S.\ Levitov, and Yu.\ V.\ Nazarov for valuable discussions and acknowledge support by the  Cambridge-MIT Institute Ltd.

\end{document}